\documentclass[english,showpacs,preprint,superscriptaddress,pop]{revtex4-1}
\usepackage{lmodern}
\usepackage{lmodern}
\usepackage[T1]{fontenc}
\usepackage{amsmath}
\usepackage{amssymb}
\usepackage{graphicx}
\usepackage{esint}

\makeatletter

\usepackage{amsthm}\usepackage{latexsym}\usepackage{bm}\usepackage{amsfonts}

\usepackage{babel}

\usepackage{hyperref}
\hypersetup{
	breaklinks = true,
    colorlinks = true,
    citecolor = {blue},
	urlcolor = {blue},
	linkcolor = {blue}
}

\usepackage{babel}

\global\long\def\EXP{\times10^}
\usepackage{cleveref}

 \global\long\def\rmd{\mathrm{d}}
 \global\long\def\rmD{\mathrm{D}}
 \global\long\def\rmc{\mathrm{c}}

 \global\long\def\bfx{\mathbf{x}}

 \global\long\def\bfv{\mathbf{v}}

 \global\long\def\bfA{\mathbf{A}}

 \global\long\def\bfB{\mathbf{B}}

 \global\long\def\bfE{\mathbf{E}}

 \global\long\def\rme{\mathrm{e}}
 \global\long\def\rmi{\mathrm{i}}

 \global\long\def\FIG#1{Fig.~\ref{#1}}
 \global\long\def\TAB#1{Tab.~\ref{#1}}

 \global\long\def\REF#1{Ref.~\cite{#1}}

 \global\long\def\calA{\mathcal{A}}

\newcommand{\WZERO}[1]{W_{\sigma_0 I}\left( #1 \right)}

\usepackage{babel}

\@ifundefined{showcaptionsetup}{}{%
 \PassOptionsToPackage{caption=false}{subfig}}
\usepackage{subfig}
\makeatother

\usepackage{babel}
\begin{document}

\title{Verification of Electromagnetic Fully-kinetic Symplectic Particle-in-cell Method in
Microinstabilities Simulation of Toroidal Plasmas}

\author{Jianyuan Xiao}

\selectlanguage{english}%

\affiliation{Department of Plasma Physics and Fusion Engineering, University of
Science and Technology of China, Hefei, 230026, China}

\author{Jian liu}
\email{Corresponding Author: liu_jian@sdu.edu.cn}

\selectlanguage{english}%

\affiliation{Weihai Institute for Interdisciplinary Research, Shandong University, Weihai 264209, China 
}
\affiliation{SDU-ANU Joint Science College, Shandong University, Weihai 264209, China}
\begin{abstract}
We present a symplectic electromagnetic fully-kinetic particle-in-cell simulation of microinstabilities in plasma, using parameters from the Cyclone Base Case [Dimits, {\sl et al.}, Physics of Plasmas 7, 969 (2000)]. The results show that the growth rates of unstable modes, including ion temperature gradient (ITG), trapped electron mode (TEM), and kinetic ballooning mode (KBM), are consistent with those obtained from previous gyrokinetic models. Additionally, the $\beta$-stabilization of the ITG is reproduced. The analysis also reveals that the impact of the ion-electron mass ratio and the numerical speed of light on the growth rate of the most unstable modes is minimal. This suggests that the fully kinetic method offers a potential for reduced computational cost when investigating the physics of drift wave instabilities at relevant space-time scales.
	
\end{abstract}



\maketitle
\section{Introduction}
It is widely believed that microinstability-driven turbulence is a key factor in the anomalous
transport \cite{horton1999drift} of plasmas, which significantly impacts the performance of 
magnetic confinement fusion devices. These instabilities arise primarily from steep temperature 
and density gradients within the plasma. Typically, a gyrokinetic 
\cite{lee1983gyrokinetic,dubin1983nonlinear,brizard2007foundations} approach is used to model this 
kind of turbulence. Previous electromagnetic gyrokinetic methods 
\cite{pueschel2010gyrokinetic,holod2013verification,pueschel2008gyrokinetic} have identified three 
types of microinstabilities at the ion motion scale, i.e., Ion Temperature Gradient (ITG), Trapped 
Electron Mode (TEM), and Kinetic Ballooning Modes (KBM) instabilities. The dominant unstable mode 
varies with plasma parameters, and simulations indicate that as plasma beta increases, these 
microinstabilities initially stabilize before becoming unstable once beta surpasses the so-called 
ballooning limit \cite{chu1978kinetic}.

Conventional gyrokinetic models simplify the evolution of both particle and electromagnetic fields 
compared to fully-kinetic charged particle-electromagnetic models, raising concerns about the 
impact of these simplifications on results. With the advent of 100 petaflop and exaflop 
supercomputers, it is now feasible to directly simulate microinstabilities in whole-volume 
magnetic fusion plasmas using fully-kinetic particle-in-cell (PIC) methods. This allows for 
verification of previous gyrokinetic models and the potential discovery of new physics that may 
not be captured by those models. However, traditional full-kinetic PIC methods face 
significant challenges when simulating charged particle-electromagnetic fields systems. First, 
the grid size must resolve the Debye radius  ($\lambda_\rmD$), such as in the Boris-Yee 
scheme, to avoid self-heating from finite-sized grid 
instabilities \cite{meyers2015numerical,barnes2021finite,ueda1994study}, which is problematic 
as the Debye radius in 
magnetic confinement fusion plasmas is typically much smaller than the global plasma size. 
For instance, with typical fusion plasma temperatures of $T_e\sim10\mathrm{keV}$ and density of
$1\EXP{20}\mathrm{m}^{-3}$, the 
Debye radius is less than $0.1\mathrm{mm}$, necessitating over 10,000 grids in each 
direction to resolve it in a typical plasma size about $1\mathrm{m}$, leading to 
computational demands 
that are currently unmanageable. Additionally, the time-step in conventional explicit PIC 
schemes is limited by the Courant-Friedrichs-Lewy (CFL) condition and plasma frequency
($\omega_\mathrm{pe}$), requiring over a million time-steps to resolve one period of
unstable modes. This long-term simulation can accumulate significant numerical 
errors, compromising result reliability. 

To address these issues, we employed a newly developed explicit 2nd-order 
charge-conservative symplectic structure-preserving electromagnetic fully-kinetic (FK) PIC scheme 
for cylindrical geometry \cite{jianyuan2021explicit,xiao2021symplectic} 
to simulate microinstabilities using the established Cyclone Base Case (CBC) \cite{dimits2000comparisons} parameters. 
This method eliminates the self-heating problem, and its superior long-term conservative 
properties ensure result accuracy. We identify frequency and growth rate of unstable modes 
and our findings include: 1) The growth rates of ITG and TEM from the FK method are slightly
higher than predictions from gyrokinetic models, while the KBM growth rate is comparable. 
2) The mode frequencies for ITG and TEM align closely with gyrokinetic predictions, whereas 
the KBM frequency is lower than expected. 3) The growth rate of the most unstable mode 
initially decreases with increasing plasma beta for ITG and TEM, while it 
significantly increases for KBM, consistent with previous electromagnetic 
gyrokinetic simulations.

This paper is organized as follows. Section 2 provides a brief introduction to the electromagnetic fully kinetic physical model and the symplectic structure-preserving method. Section 3 details the construction of Tokamak equilibrium, including Cyclone Base Case parameters and simulation results. Finally, Section 4 concludes the paper.

\section{Symplectic Electromagnetic fully-kinetic particle-in-cell scheme employed in SymPIC}
First, we briefly introduce the symplectic electromagnetic fully-kinetic PIC scheme
in the cylindrical coordinate developed previously \cite{jianyuan2021explicit} which is implemented
in SymPIC code and it can run efficiently on modern heterogeneous 
supercomputers with up to over 40 million cores \cite{xiao2021symplectic}. Previously, we have 
applied it to simulate Tokamak edge plasmas and obtained comparable
results with fluid simulations and experiments \cite{liu2022experimental}. In this work, we focus
on verifying of simulations of microinstabilties in core plasmas using this approach.

This work focuses on the physical model of charged particles coupled with electromagnetic 
fields, of which the evolution equations are
\begin{eqnarray}
	\ddot{\bfx}_{s,p}&=& \frac{q_s}{m_s}\left( \bfE\left( \bfx_{s,p} \right)+\dot\bfx_{s,p}\times\bfB\left( \bfx_{s,p} \right) \right)~,\label{EqnLRZ}\\
	\dot\bfE&=& \nabla\times\bfB-\sum_{s,p}q_s\dot{\bfx}_{s,p}\delta\left( \bfx-\bfx_{s,p} \right)~,\\
	\dot{\bfB}&=& -\nabla\times\bfE~,\label{EqnDBDT}
\end{eqnarray}
where $\bfx_{s,p}, m_s, q_s$ indicates the location, mass, charge of the $p$-th particle of 
the $s$-species, $\bfE$ and $\bfB$ are electromagnetic fields. For simplicity,
we normalize both permittivity $\epsilon_{0}$ and permeability $\mu_{0}$
in the vacuum to 1. To build symplectic structure preserving PIC method in the cylindical 
geometry, we start from the action integral of the charged particle coupled with electromagnetic
fields system, which reads
\begin{eqnarray}
\calA[\bfx_{sp},\dot{\bfx}_{sp},\bfA,\phi] & = & \int\rmd t\sum_{s,p}\left(L_{sp}\left(m_{s},\bfv_{sp}\right)+q_{s}\left(\bfv_{sp}\cdot\bfA\left(\bfx_{sp},t\right)-\phi\left(\bfx_{sp},t\right)\right)\right)\nonumber \\
 &  & +\int\rmd V\rmd t\frac{1}{2}\left(\left(-\dot{\bfA}\left(\bfx,t\right)-\nabla\phi\left(\bfx,t\right)\right)^{2}-\left(\nabla\times\bfA\left(\bfx,t\right)\right)^{2}\right)~,
\end{eqnarray}
where $\bfx=[x_{1},x_{2},x_{3}]$ and 
$\rmd V=|h_{1}\left(\bfx\right)h_{2}\left(\bfx\right)h_{3}\left(\bfx\right)|\rmd x_{1}\rmd x_{2}\rmd x_{3}$. In cylindrical coordinate, we choose line elements as 
\begin{eqnarray}
	h_1\left( x_1,x_2,x_3 \right)&=& 1~,\\
	h_2\left( x_1,x_2,x_3 \right)&=& (x_1/X_0+1)\Delta x_2/\Delta x_1~,\\
	h_3\left( x_1,x_2,x_3 \right)&=& \Delta x_3/\Delta x_1~.
\end{eqnarray}
Here $\Delta x_1, \Delta x_2 $ and $\Delta x_3$ are reference grid size for each direction, $X_0$ is 
the location of inner boundary of the simulation domain. Hence the 
velocity of each particle is
\begin{eqnarray}
	\bfv_{sp}&=& \left[ \dot x_{1,sp}h_1,\dot x_{2,sp}h_2,\dot x_{3,sp}h_3 \right]~,
\end{eqnarray}
where
\begin{eqnarray}
	\left[ x_{1,sp},x_{2,sp}, x_{3,sp} \right]=\bfx_{sp}
\end{eqnarray}
is the coordinate of each particle in the cylindrical geometry. Finally the Lagrangian for each
particle is 
\begin{eqnarray}
	L_{sp}\left(m_{s},\bfv_{sp}\right)&=& \frac{1}{2}m_s\bfv_{sp}^2,
\end{eqnarray}
and the electromagnetic fields are defined as 
\begin{eqnarray}
	\bfE&=& -\dot{\bfA}-\nabla\phi~,\\
	\bfB&=& \nabla\times\bfA~.
\end{eqnarray}
The evolution equations Eqs. (\ref{EqnLRZ})-(\ref{EqnDBDT}) can be derived from
variation, i.e.,
\begin{eqnarray}
	\frac{\delta \calA}{\delta \bfx_{sp}}&=& 0~,\\
	\frac{\delta \calA}{\delta \bfA}&=& 0~,\\
	\frac{\delta \calA}{\delta \phi}&=& 0~,
\end{eqnarray}
and the key to build symplectic algorithm is to find the discrete action integral.
According to our previous 
work \cite{xiao2021symplectic}, the action integral can be discretized using Discrete 
Exterior Calculus (DEC), Whitney interpolating form and zigzag integral path of particles
to guarantee the discrete gauge invariance property. In such discretization, local charge 
conservation property is automatically satisfied and hence we can directly use electromagnetic
fields rather than potentials to build the algorithm.
Then the final explicit iteration scheme can be obtained by using discrete variational method
or Hamiltonian splitting technique. Detail of the 2nd-order charge-conservative symplectic 
PIC scheme for the cylindrical coordinate can be found in the Appendix B of 
\REF{xiao2021symplectic}. In the present work, the Whitney interpolating 0-form in slab geometry is choosen as
\begin{eqnarray}
	\WZERO{\bfx}=W_1\left( x_1 \right)W_1\left( x_2 \right)W_1\left( x_3 \right)~,
\end{eqnarray}
where
\begin{equation}
	W_1\left( x \right)=\left\{
	\begin{array}{lc}
		0,&x<-2\\
		\frac{\left( x+2 \right)^2}{4},&-2\leq x < -1\\
		\frac{2-x^2}{4},&-1\leq x<1\\
		\frac{\left( x-2 \right)^2}{4},&1\leq x < 2\\
		0,&x\geq 2~.
	\end{array}
	\right.
\end{equation}

The present method offers several advantages over previous gyrokinetic methods. 1)
It treats charged particles as fully kinetic, avoiding the gyro-average or guiding
center approximation of charged particles. 2) The model of fields
is complete electromagnetic, incorporating electromagnetic waves, unlike previous 
gyrokinetic methods that ignore the 
displacement current term $\partial \bfE/\partial t$ in the Maxwell equation. 
3) The model includes both the parallel and perpendicular components of the 
magnetic potential, accounting for the effects of both shear Alfven 
and compressible magnetosonic waves, which aro often simplified in earlier electromagnetic 
gyrokinetic models. Due to the computational cost
constraints, using a reduced speed of light in vacuum and a reduced ion-electron
mass ratio is preferred for real simulations. However, the impact of these reductions 
on the results remains uncertain and will be explored in this study.

\section{Simulation of instabilities in core plasmas}

The Cyclone Base Case (CBC) parameters, typically used to benchmark gyrokinetic algorithms,
\cite{dimits2000comparisons,rewoldt2007linear,garbet2010gyrokinetic} 
characterize a local Tokamak plasma. They can be listed as follows. 
The minor radius $r=0.5a$, where $a$ is the minor radius of the last closed flux surface.
The shape of poloidal cross section is circular, with $n_i=n_e$, $T_i=T_e$, 
where $n_i$, $n_e$, $T_i$, $T_e$ are the ion density, electron density, 
ion temperature and electron temperature, respectively. The parameter values in dimensionless
form are $\eta_i= L_\mathrm{n}/L_\mathrm{T}=3.114$, where $L_\mathrm{n}$ and $L_\mathrm{T}$ 
are the density and temperature gradient scale lengths, respectively,
safety factor which is choosen as $q=rB_\mathrm{t}/R_0B_\mathrm{p}=1.4$, where $R_0$ is the
major radius and $B_\mathrm{t}$ and $B_\mathrm{p}$ are the toroidal and poloidal
magnetic field components, $\hat{s}=(r/q)dq/dr=0.8$, 
$R_0/L_\mathrm{T}=6.92$, and $a/R_0=0.36$.
We build initial conditions for kinetic particles and 
magnetic field from the 2D solution that satisfing the above local conditions 
generated by a Free boundary Grad-Shafranov solver (FreeGS) \cite{jeon2015development}, 
then perform a 2D-3V (i.e., $N_\theta=1$) simulation using SymPIC to generate 
a relatively stable kinetic equilibrium as
the final initial condition for 3D-3V simulation.
The gyro-radius of thermal speed
for ions at $r$ is set to $\rho^*=m_i\sqrt{T_i/m_i}/(q_iB_0)\sim a/147$, and we tested various
plasma betas of electrons, i.e., $\beta_\rme=0.17\%,0.58\%,0.87\%,1.16\%$ and $1.45\%$. 
According to previous 
results based on gyrokinetic simulations,
\cite{pueschel2010gyrokinetic,holod2013verification,pueschel2008gyrokinetic}
growth rate of microinstabilties such as ion temperature gradient (ITG) and trapped 
electron modes (TEM) will become small with the increment of plasma beta. Simulation domain
is a 1/3 ring, i.e., $R_{\mathrm{left}}\leq R<R_{\mathrm{left}}+1.033R_0, 0\leq\theta<2\pi/3, 0<z<1.381R_0$, 
$R_0=0.958\mathrm{m}$, $R_{\mathrm{left}}= 0.437 R_0$,
where
\begin{eqnarray}
	R&=& x_1\Delta x_1+R_{\mathrm{left}}~,\\
	\theta&=& x_2\Delta x_2/R_{\mathrm{left}}~,\\
	z&=& x_3\Delta x_3~,\\
	\Delta x_1&=& 1.033R_0/N_R~,\\
	\Delta x_2&=& \left( 2\pi/3 \right) R_{\mathrm{left}}/N_\theta~,\\
	\Delta x_3&=& 1.33\Delta x_1~,
\end{eqnarray}
and number of 
grids in $R, \theta, z$ directions $N_R, N_z $ and $N_\theta$ are set
to $N_R=N_z=256, N_\theta=64$. Timestep is set to $2.02\EXP{-3}R_0/\rmc_N$ where $\rmc_N$ is 
the numerical speed of light in vacuum in the simulations.
To reduce the computational complexity, $\rmc_N$ 
is set to $0.225\rmc$ where $\rmc$ is real speed of light in vacuum, mass of electron is set 
to $m_e=m_i/100$, ions are all deterium. 
Thermal speeds and densities of electrons and ions at the magnetic axis 
are set to $v_\mathrm{te}=10v_\mathrm{ti}=0.0447\rmc_N$ and 
$n_\rme=n_\rmi=2.65\EXP{21}\beta_\rme\mathrm{m}^{-3}$, respectively. Total number of time-steps
is $N_t=2\EXP{6}$, which is about $N_t\Delta t\sim4.7\EXP{3}\omega_\mathrm{ci}^{-1}\sim18R_0/v_\mathrm{ti}$. The maximum number of sampling particle in one grid cell for each species ranges from 
$1000-4000$ dependending on different parameters. One above 3D-3V simulation requires
approximately $1\EXP{5}$ to $4\EXP{5}$ core hours CPU time on the new Sunway supercomputer.
\begin{figure}[!t]
	\begin{center}
		\includegraphics[width=0.7\linewidth]{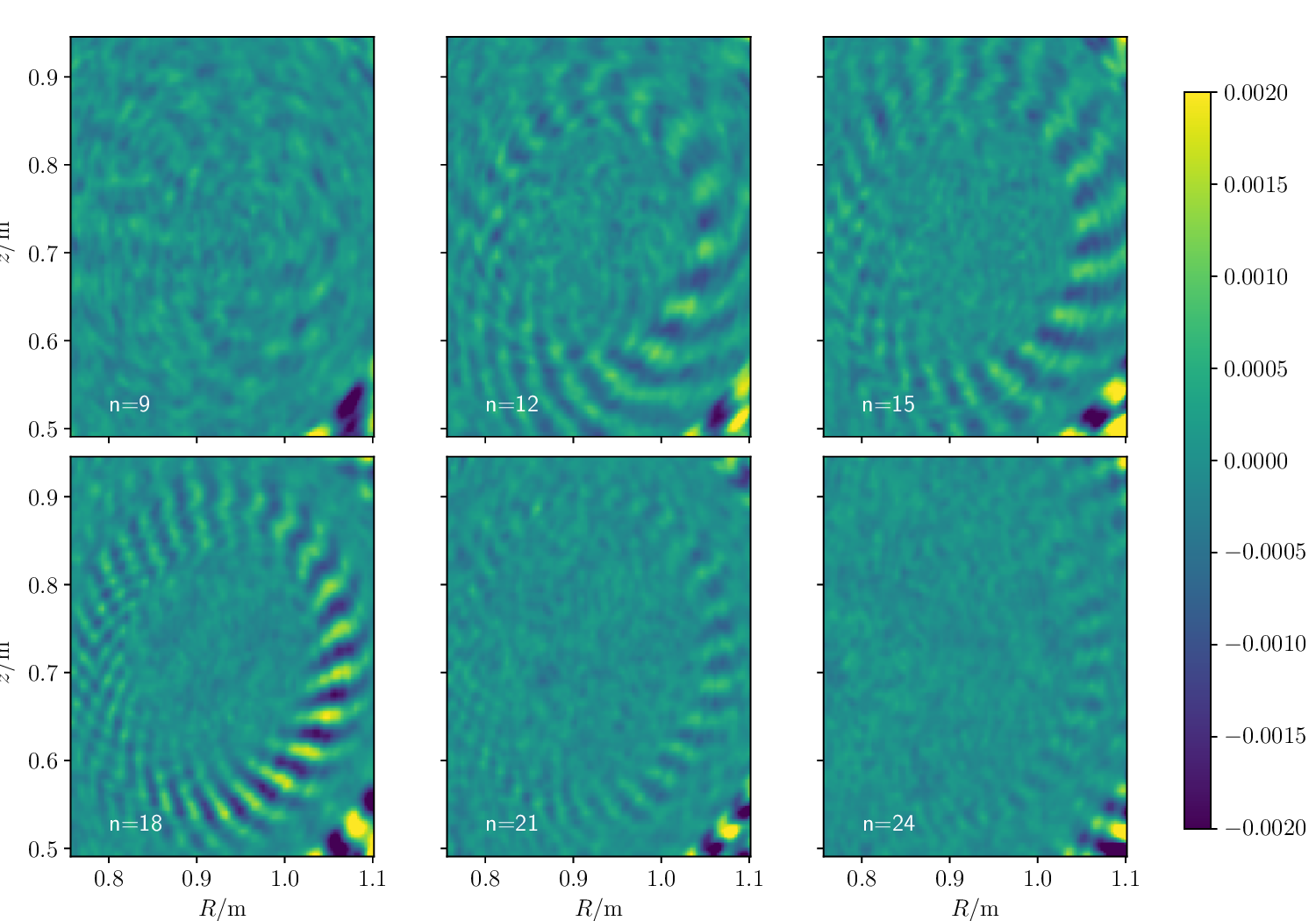}
	\end{center}
	\caption{Unstable mode structures of density perturbation for different toroidal mode number $n$ in CBC-like plasma with ITG parameters, here $tv_\mathrm{ti}/R_0\sim10.35$ and $\beta_\rme=0.17\%$.}
	\label{FigModITG}
\end{figure}
\begin{figure}[!t]
	\begin{center}
		\includegraphics[width=0.7\linewidth]{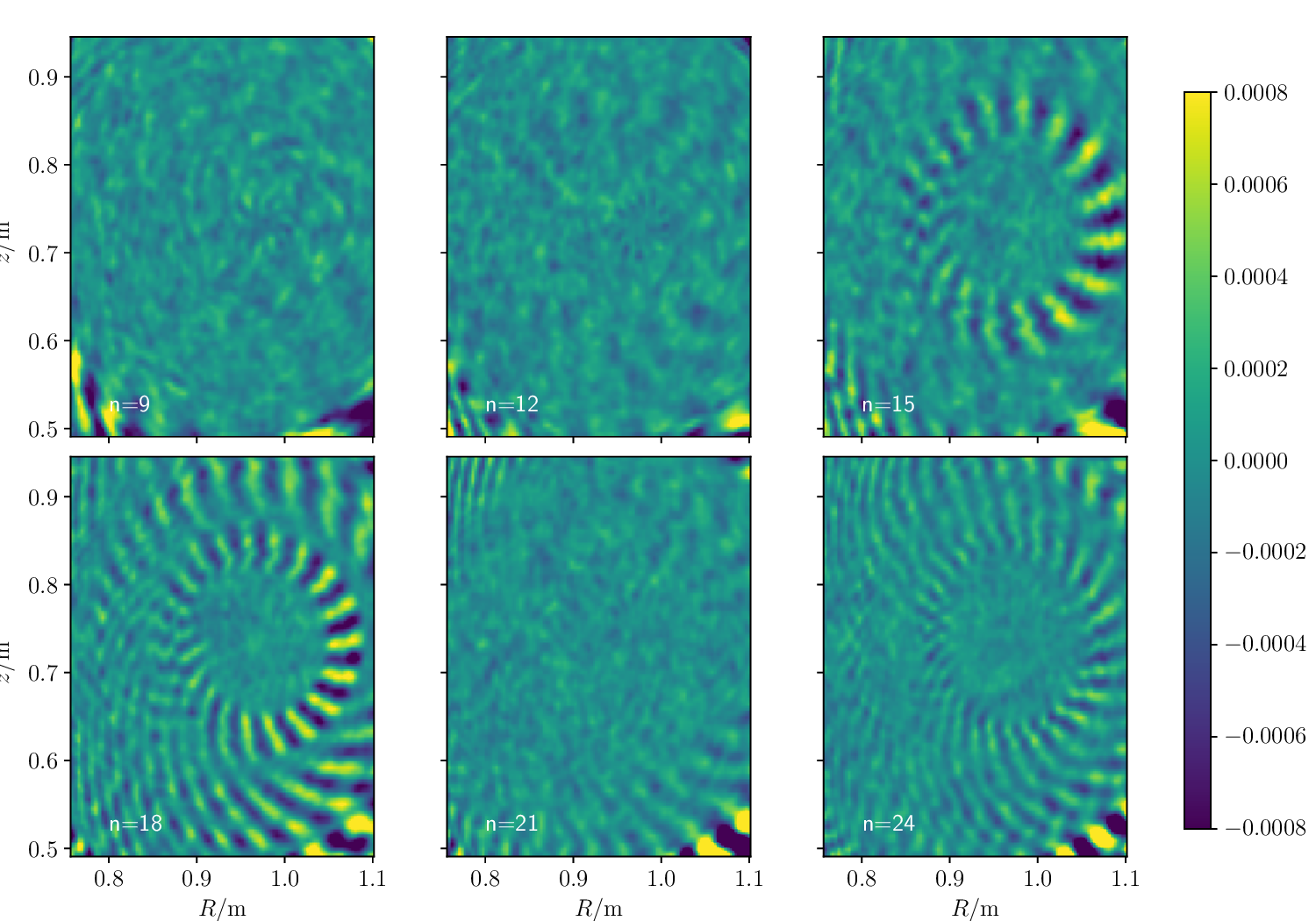}
	\end{center}
	\caption{Unstable mode structures of density perturbation for different toroidal mode number $n$ in CBC-like plasma with TEM parameters, here $tv_\mathrm{ti}/R_0\sim9.35$ and $\beta_\rme=1.16\%$.}
	\label{FigModTEM}
\end{figure}
\begin{figure}[!t]
	\begin{center}
		\includegraphics[width=0.7\linewidth]{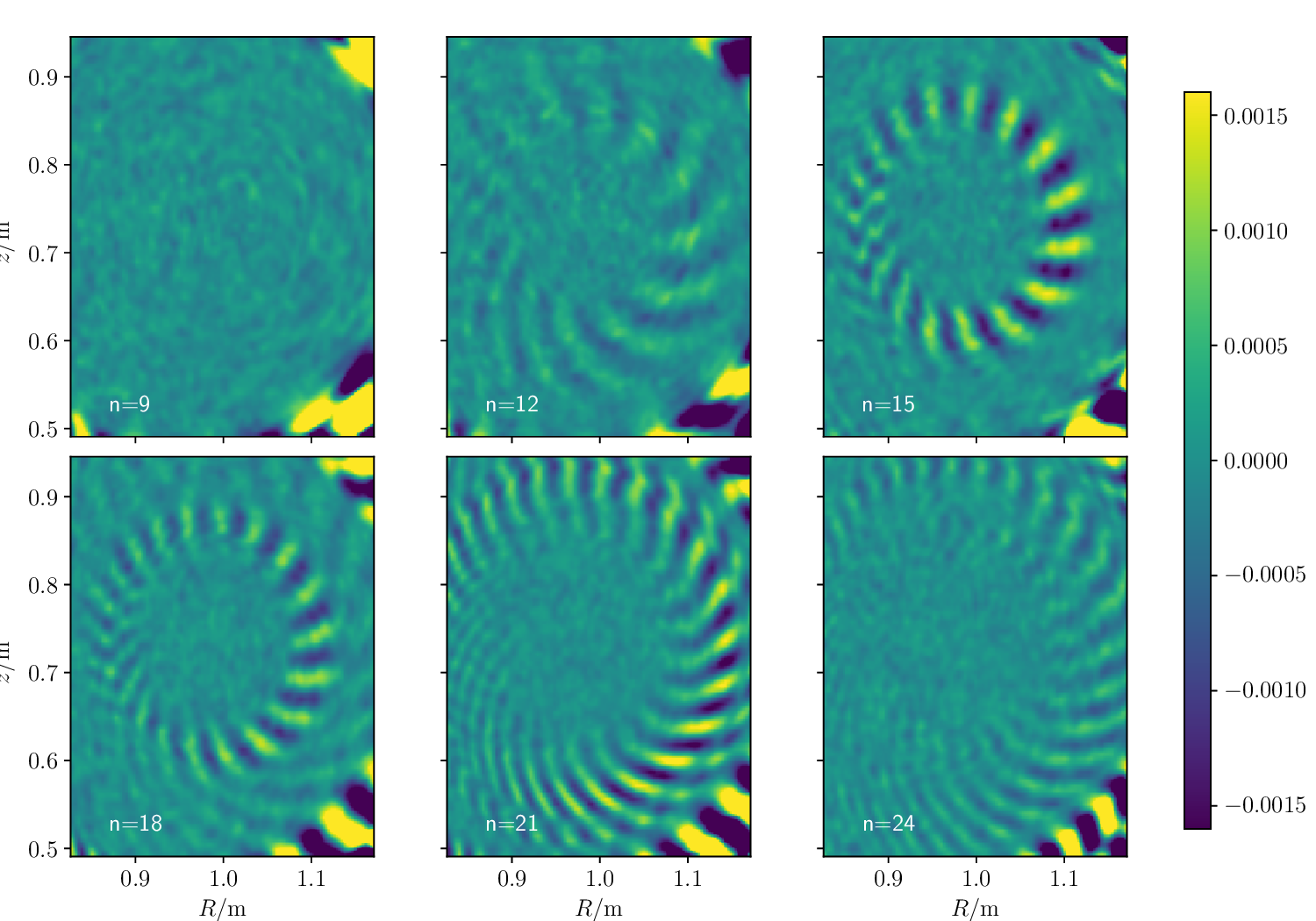}
	\end{center}
	\caption{Unstable mode structures of density perturbation for different toroidal mode number $n$ in CBC-like plasma with KBM parameters, here $tv_\mathrm{ti}/R_0\sim8.76$ and $\beta_\rme=1.45\%$.}
	\label{FigModKBM}
\end{figure}

\begin{figure}[htp]
	\begin{center}
		\includegraphics[width=0.65\linewidth]{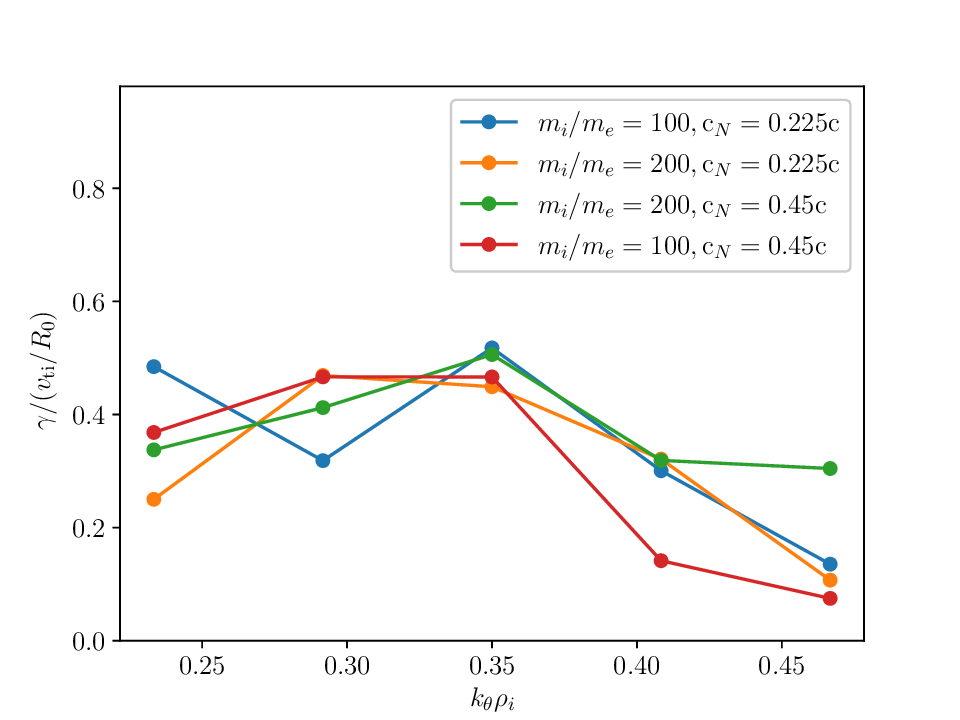}
	\end{center}
	\caption{Dependence of the growth rate $\gamma$ on the $k_\theta$ obtained by SymPIC for $m_\rmi=200m_\rme$ and $m_\rmi=100m_\rme$}
	\label{FigMIMERAT}
\end{figure}
\begin{figure}[htp]
	\begin{center}
		\includegraphics[width=0.6\linewidth]{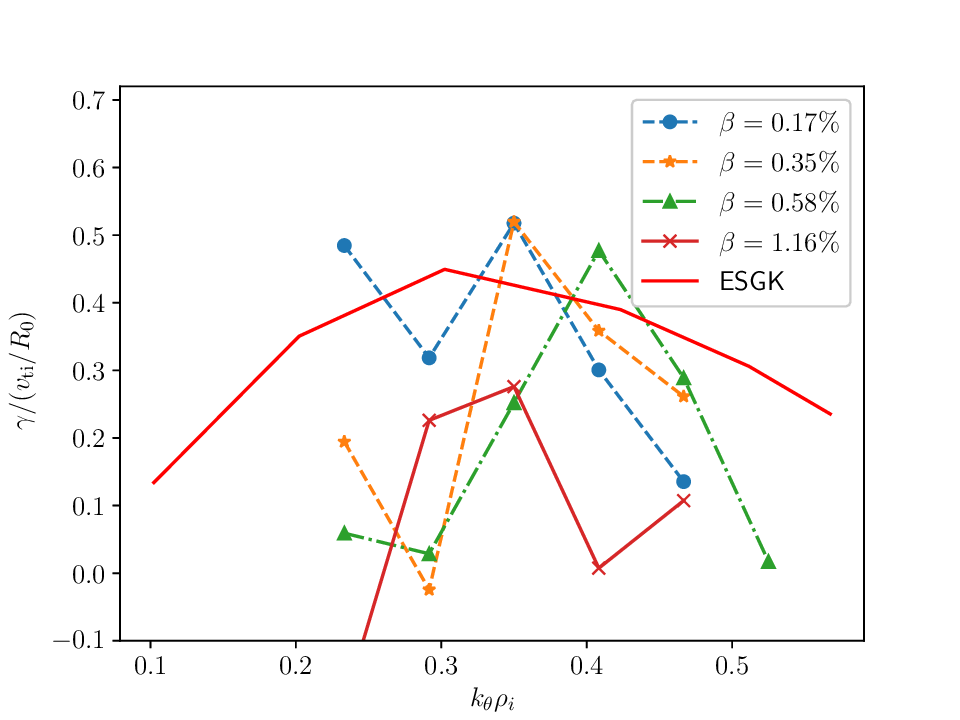}
	\end{center}
	\caption{Dependence of the growth rate $\gamma$ on the $k_\theta$ obtained by SymPIC, ESGK ($\beta_\rme=0$)
	results \cite{rewoldt2007linear} are also shown as reference.}
	\label{FigIncRate}
\end{figure}
\begin{figure}[htp]
	\subfloat[Frequence.]{\includegraphics[width=0.49\textwidth]{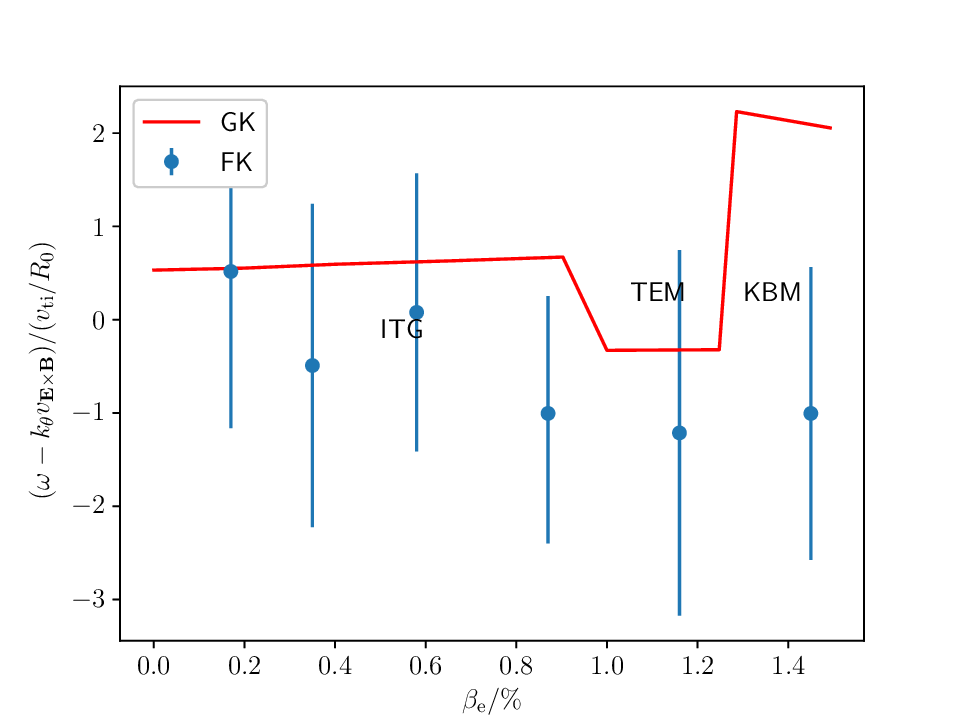}\label{FigFreq}}
	\subfloat[Growth rate.]{\includegraphics[width=0.49\textwidth]{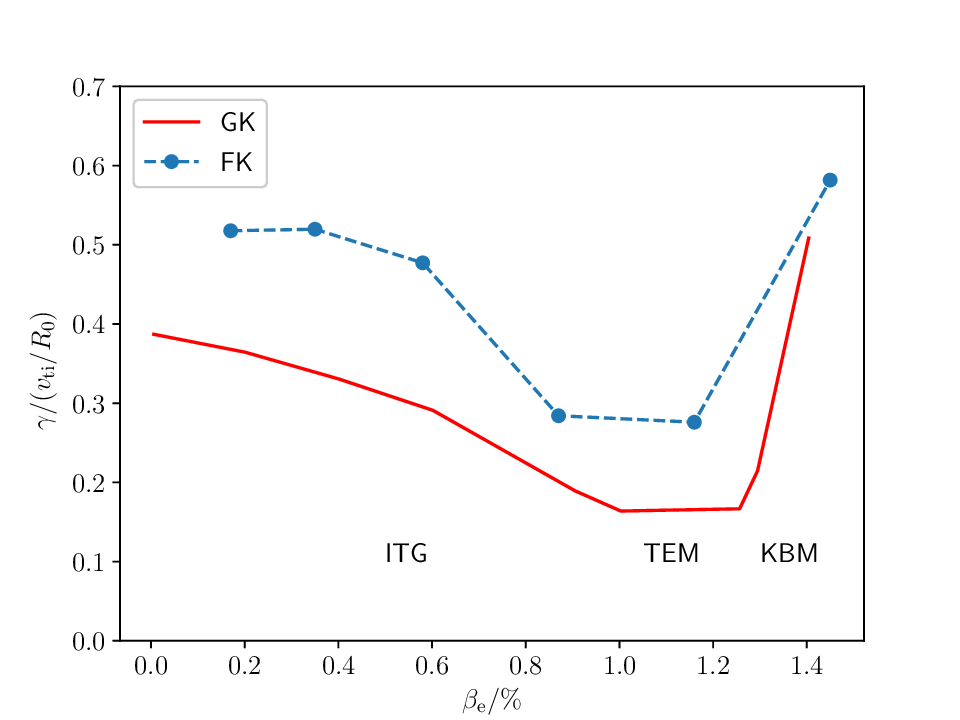}}
	\caption{Dependence of frequency $\omega-k_\theta v_{\mathbf{E\times B}}$ and growth rate $\gamma$ on the plasma $\beta$ obtained by SymPIC. Electromagnetic GK results \cite{pueschel2008gyrokinetic} are also shown as reference.}
	\label{FigIncRateBeta}
\end{figure}

Structures of unstable modes for $\beta_\rme=0.17\%, 1.16\%$ and $1.45\%$ are shown in Figs. \ref{FigModITG}-\ref{FigModKBM}. For $\beta_\rme=1.45\%$ case, due 
to the large Shafranov shift, we modify $R_{\mathrm{left}}$ as $R_{\mathrm{left}}=0.509R_0$, corresponding new major radius is $R_{0,\mathrm{KBM}}=1.06R_0$.
mode number of the most unstable mode is $n=18$, so the corresponding $k_\theta=nq/r$ complies
$k_\theta\rho_i=0.34$
which is consistent with previous electrostatic gyrokinetic (ESGK) 
simulations \cite{rewoldt2007linear}. Here in our simulation all toroidal modes are nonlinearly 
coupled together, so the growth rate of unstable toroidal modes may different from that 
obtained by conventional methods in which only one toroidal mode is included. 
To see how the ion-electron mass ratio and numerical speed of light effect the results,
we also perform 3 extra cases with the changing of parameters listed in \TAB{TabPARAM},
\begin{table}
	\centering
	\begin{tabular}{|c|c|c|c|}\hline
		Extra Case No.&1&2&3\\\hline
		$m_i/m_e$&200&200&100\\\hline
		$\mathrm \rmc_N/\rmc$&0.225&0.45&0.45\\\hline
	\end{tabular}
	\caption{Parameters for testing the dependence of growth rates to the ion-electron mass ratio and the numerical speed of light.}
	\label{TabPARAM}
\end{table}
and other parameters remains the same as 
the $\beta_\rme=0.18\%$ case. We plotted growth rates and frequencies 
of unstable modes obtained by
SymPIC for these cases in \FIG{FigMIMERAT}, which 
shows that the growth rate of the most unstable mode 
is relatively independent from the ion-electron mass ratio and the numerical 
speed of light.
The growth rate of 
unstable modes of plasmas with different 
$\beta_\rme$ and $k_\theta$ are presented in \FIG{FigIncRate}, 
we also show results obtained by 
previous ESGK ($\beta_\rme=0$) simulations for comparison. It can be seen that 
when $\beta_\rme$ is low 
($\beta_\rme=0.18\%$), growth rates obtained from electromagnetic fully-kinetic PIC scheme 
are in the same order of magnitude and roughly have the same trend with changes in $k_\theta$. 
When the pressure becomes high, the growth rate drops. To perform a more comprehensive 
comparison, frequency and growth rate of the most unstable mode obtained by 
the present fully-kinetic (FK) PIC method are plotted
in \FIG{FigIncRateBeta}. Electromagnetic gyrokinetic (GK) simulation
results \cite{pueschel2008gyrokinetic} are also plotted as reference. It is clear that the 
growth rate obtained by SymPIC and previous gyrokinetic methods are in the same order of 
magnitude and have the same trend with changes in $\beta_\rme$.

The mode frequencies obtained
by the FK method has large uncertainty (see the errorbars in \FIG{FigFreq}) due to the large
noise of radial electric field in the 
PIC simulation. However they are also in the same order of magnitude compared with corresponding
GK method, expect for the KBM case. This may due to that converting the KBM frequency from 
the laboratory frame to the plasma frame requires no or more than subtraction of
the $\bfE \times \bfB$ velocity. 

\section{Discussions and Conclusions}
In this study, we have presented a comprehensive investigation of microinstabilities in 
toroidal plasmas using a novel symplectic electromagnetic FK PIC simulation method. 
By employing the well-established CBC parameters, we have successfully verified the reliability 
of our fully-kinetic approach in capturing the growth rates and mode frequencies of key 
microinstabilities, including ITG, TEM, and KBM instabilities. The growth rate of the unstable mode obtained through the FK approach slightly differs from, yet remains within the same order 
of magnitude as, that derived from previous GK methods. 
These results thereby providing a robust benchmark for future studies. 

The FK PIC method employed in this work offers several significant advantages over 
traditional gyrokinetic approaches. It fully accounts for the kinetic nature of charged 
particles without relying on gyro-average or guiding center approximations. Additionally,
it incorporates the complete electromagnetic field model, including the displacement current 
term often neglected in gyrokinetic simulations. This treatment allows for a more accurate 
representation of the underlying physics, especially in scenarios where electromagnetic
effects and high-frequency phenomena play crucial roles.

Our simulations reveal that the growth rates of the most unstable modes are relatively 
insensitive to variations in the ion-electron mass ratio and the numerical speed of light. 
This finding suggests that the fully-kinetic method has the potential to significantly 
reduce computational costs when investigating drift wave instabilities at relevant 
space-time scales, making it a promising tool for large-scale simulations of magnetic
fusion plasmas.

The results also highlight the $\beta$-stabilization effect on ITG instabilities and the 
contrasting behavior of KBM instabilities as plasma $\beta$ increases. These observations 
align well with previous electromagnetic gyrokinetic simulations, further validating 
the accuracy and applicability of our fully-kinetic approach.

In conclusion, this study underscores the importance of employing fully-kinetic 
simulations to gain deeper insights into the complex dynamics of microinstabilities in toroidal
plasmas. The successful implementation of the SymPIC method on modern heterogeneous supercomputers 
demonstrates its potential for high-performance computing applications. Future work may focus on 
extending this approach to more complex plasma configurations and exploring additional physics 
phenomena that could be uncovered through fully-kinetic simulations. Overall, the findings 
presented here pave the way for more accurate and efficient modeling of plasma microinstabilities, 
ultimately contributing to the advancement of magnetic fusion research.

\begin{acknowledgments}
This work was supported by the Strategic Priority Research Program of
Chinese Academy of Sciences (Grant No. XDB0500302), the National MC Energy R\&D Program
\\(2024YFE03020004, 2018YFE0304100), and the National Natural Science
Foundation of China (NSFC-11905220 and 11805273). J. Xiao would like to thank Prof. Jinlin Xie and 
Prof. Weixing Ding at University of Science and Technology of China for valuable discussions on 
physics of drift wave instabilities.
Simulations are performed in the 
New Sunway supercomputer, Tianhe-3A prototype machine, Hanhai20 of the USTC.
\end{acknowledgments}

\bibliographystyle{apsrev4-1}
\bibliography{verify_fk}

\end{document}